\documentclass[universe,article,accept,pdftex,moreauthors]{Definitions/mdpi} 

\firstpage{1} 
\pubvolume{1}
\issuenum{1}
\articlenumber{0}
\pubyear{2023}
\copyrightyear{2023}
\datereceived{ } 
\daterevised{ } % Comment out if no revised date
\dateaccepted{ } 
\datepublished{ } 
\hreflink{https://doi.org/} % If needed use \linebreak
\Title{The dead cone effect in heavy quark jets observed in momentum space and its QCD explanation}

\TitleCitation{The dead cone effect in heavy quark jets observed in momentum space and its QCD explanation}

\Author{Stefan Kluth$^{1,*}$, Wolfgang Ochs$^{1}$ and Redamy Perez Ramos$^{2}$}

\AuthorNames{Stefan Kluth, Redamy Perez Ramos and Wolfgang Ochs}

\AuthorCitation{Kluth, S.; Ochs, W.; Perez Ramos, R.}
\address{%
$^{1}$ \quad Max-Planck-Institut f\"ur Physik, Boltzmannstr. 8, 85748 Garching, Germany\\
$^{2}$ \quad DRII-IPSA, Bis, 63 Boulevard de Brandebourg, 94200 Ivry-sur-Seine, France; LPTHE, UMR 7589, Sorbonne Universit\'e et CNRS, 4 place Jussieu, 75252 Paris Cedex 05, France}

\corres{Correspondence (speaker): skluth@mpp.mpg.de}

\firstnote{MPI f\"ur Physik preprint number: MPP-2023-293}

\abstract{The production of a heavy quark is accompanied by gluon bremsstrahlung with angular and momentum spectra predicted by perturbative Quantum Chromo Dynamics (QCD). The radiation off heavy quarks is predicted to be suppressed for large momentum particles, as a consequence of the angular ``dead cone effect''. In this paper, we studied this effect using data from Z boson decays to c- or b-quarks in $e^+e^-$ annihilation. The momentum spectra for charged particles are reconstructed in the momentum fraction variable $x$ or $\xi=\ln(1/x)$ by removing the decays of the heavy hadrons. We find an increasing suppression of particles with rising $x$ down to a fraction of $\lesssim 1/10$ for particles with $x\gtrsim0.2$ in b-quark and $x\gtrsim0.4$ in c-quark jets in comparison to light quark momentum spectra. The sensitivity to the dead cone effect in the present momentum analysis is larger than in the recently presented angular analysis. The suppression  for c- and b-quark fragmentation is in good quantitative agreement with the expectations based on perturbative QCD within the Modified Leading Logarithmic Approximation (MLLA) in the central kinematic region. The data also support a two parameter description in the MLLA of these phenomena (``Limiting Spectrum''). The sensitivity of these measurements to the heavy quark mass is investigated.}


\begin{document}

%%%%%%%%%%%%%%%%%%%%%%%%%%%%%%%%%%%%%%%%%%
% \setcounter{section}{-1} %% Remove this when starting to work on the template.

\section{Introduction}

We summarise here the work published in~\cite{Kluth:2023umf}. The dead cone effect is a prediction of QCD, the theory of strong interactions within the Standard Model of particle physics. It originates from the radiation pattern off a heavy quark as obtained in perturbation theory~\cite{Dokshitzer:1991fc,Dokshitzer:1991fd}. For an energetic heavy quark $Q$ of mass $M_Q$ and energy $E_Q$ such that $E_Q/M_Q \gg 1$, the gluon emission probability for small emission angle $\Theta$ and low energy $\omega$ can be written as
\begin{equation}
  d\sigma_{Q\to Q+g} \simeq \frac{\alpha_S}{\pi}C_F\frac{\Theta^2d\Theta^2}{(\Theta^2+\Theta^2_0)^2} \frac{d\omega}{\omega},
\label{emission}
\end{equation}
with angular cut-off $\Theta_0=M_Q/E_Q$, $\alpha_S$ is the strong coupling constant and $C_F=4/3$ is the QCD colour factor. Therefore, for smaller emission angles $\Theta<\Theta_0$, gluon radiation is suppressed and vanishes in a cone (called ``dead cone'') around the flight direction of the heavy quark $Q$. For large emission angles $\Theta\gg \Theta_0$, the gluon radiation pattern becomes identical to that of a light quark jets. Since for gluon radiation small emission angles and large momenta are correlated large momentum radiation is expected to be suppressed as well.

An observation of the dead cone effect has been reported by the ALICE collaboration~\cite{ALICE:2021aqk}. The angular spectra of subjets within charm-tagged jets and inclusive jets were studied in proton-proton collisions at $\sqrt{s}=13$~TeV at the Large Hadron Collider (LHC). A relative suppression of small angle subjet emission is observed in the heavy quark jet in agreement with Monte Carlo Event Generators (MCEG) combining the hard interactions of the partons from the protons with a QCD parton shower and a hadronisation model. The suppression is most significant for energies of the branching subjets $E_{radiator} < 10$~GeV. 

In~\cite{Kluth:2023umf} the idea is to study the momentum spectra of prompt charged particles in heavy flavour Q (b or c) tagged jets produced in $e^+e^-$ annihilation at centre of mass (cms) energies $\sqrt{s}=m_Z$ corresponding to the Z boson peak. Prompt charged particles refer to those produced by parton shower like radiation off the heavy quark and excludes particles produced in decays of heavy hadrons. Predictions of QCD in the modified leading log approximation (MLLA) for these momentum spectra were presented in~\cite{Dokshitzer:1991fc,Dokshitzer:1991fd} and shown to be a direct consequence of the dead cone effect in momentum space after integrating over the angular variable. The energies of the heavy quark jets are $E_{jet}\simeq 45$~GeV for Z boson decays to a pair of heavy quarks.

\section{Results}

The so-called fragmentation function $\bar{D}(\xi,W)$ for charged particles $h$ with momentum fraction $x=2p/\sqrt{s}$ and $\xi=\log(1/x)$ at reference energy in $e^+e^-$ annihilation $\sqrt{s}=W$ is given as the single inclusive cross section:
\begin{equation}\label{eq:FF}
  \bar{D}(\xi,W) = \frac{1}{2}\frac{1}{\sigma_{tot}} \frac{d\sigma^h}{d\xi} (\xi,W)\;\;.
\end{equation}
The $\xi$-distributions of events with heavy flavour (Q=b,c) tagged jets including the heavy hadron decay products have been measured by several groups~\cite{DELPHI:1998cgx,OPAL:1998arz}. 

In order to extract the fragmentation functions of heavy quarks $\bar{D}_Q(\xi,W)$ the contributions to the measured momentum spectra from heavy hadron decay 
products have to be subtracted. The momentum spectra of charged B-hadron or Charm-hadron decay products have not been measured separately\footnote{A measurement 
could use the impact parameters of reconstructed tracks w.r.t.\ the primary vertex to select the heavy hadron decay products.}. In the analysis these spectra are 
calculated with the Monte Carlo event generator (MCEG) program Pythia8~\cite{Bierlich:2022pfr,Skands:2014pea} based on samples of 100'000 simulated Z boson decays 
to heavy quarks. The B hadron decay multiplicities in Z boson decays to heavy quarks have been measured precisely~\cite{Dokshitzer:2005ri}. The simulated $\xi$ 
spectrum for charged B hadron decay products is rescaled by 13\% to match the observed multiplicity while the corresponding distribution for Charm hadron decays does not need rescaling. 

\begin{figure}[H]
\begin{tabular}{cc}
\includegraphics[width=0.475\linewidth]{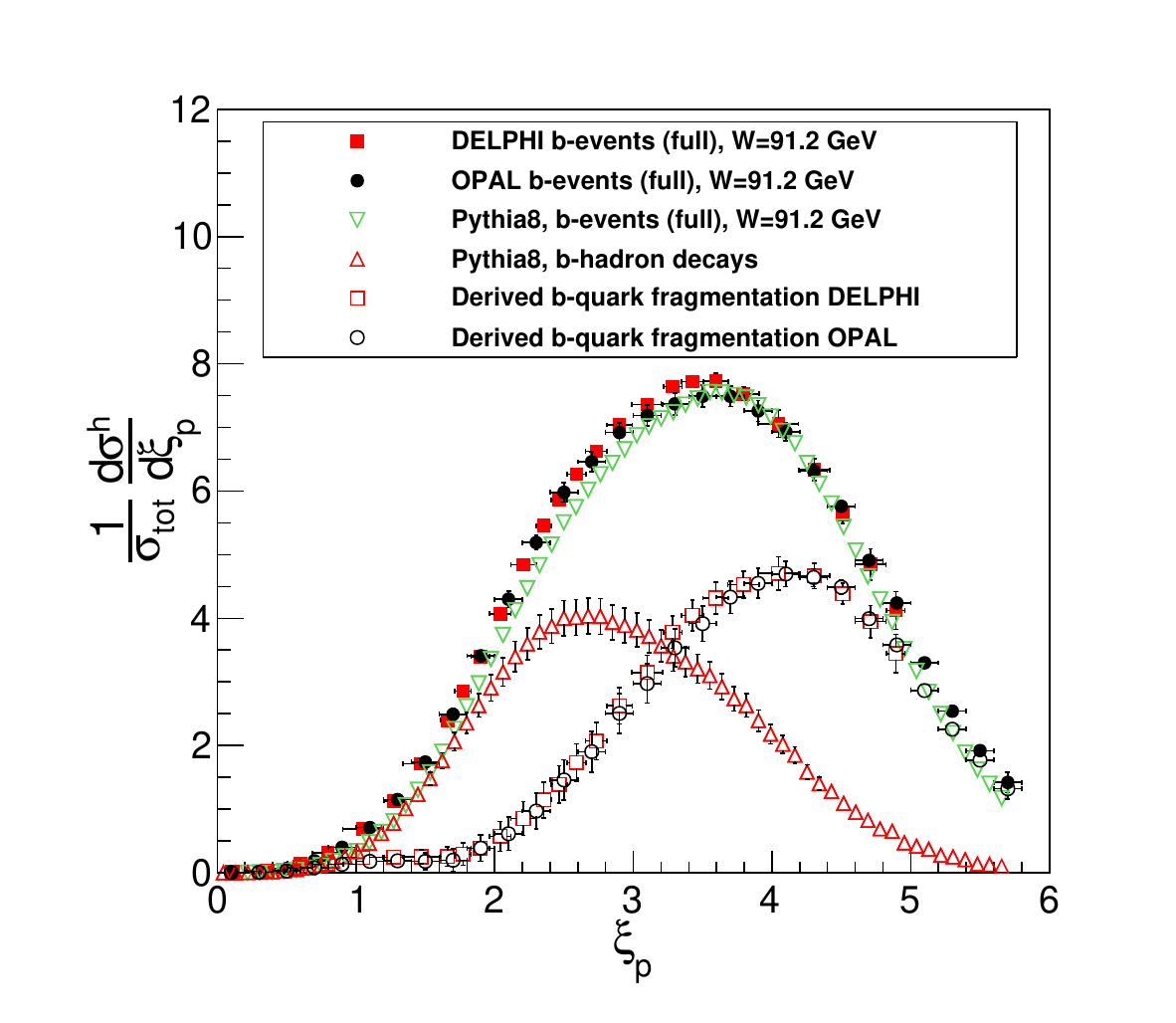} &
\includegraphics[width=0.5\linewidth]{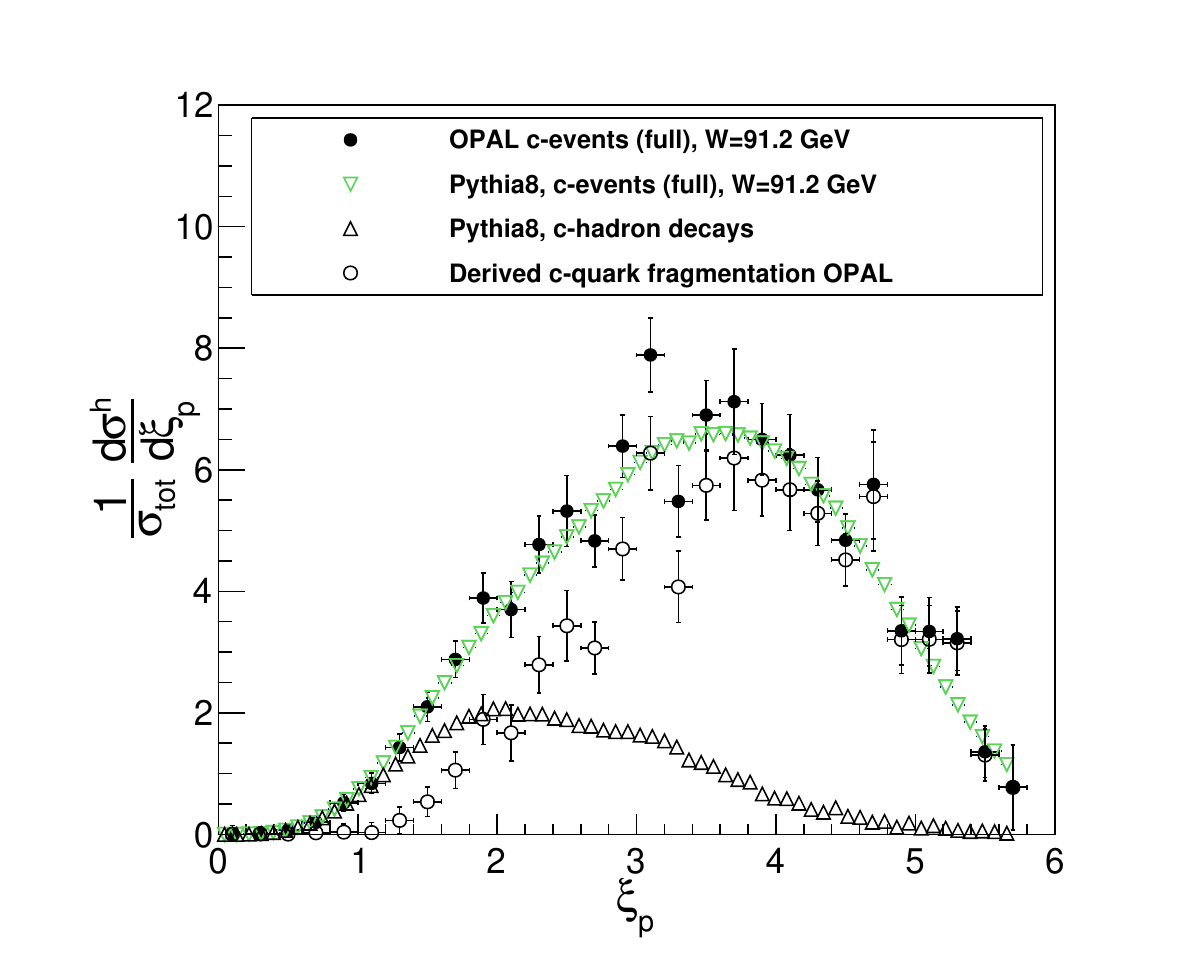} \\
\end{tabular}
\caption{(left) Momentum spectrum of charged hadrons in $\xi=\log(1/x)$ for Z decays to $b\bar b$ including B-hadron decays at 91.2~GeV from DELPHI and OPAL and the prediction from the Pythia8 MCEG. The b-quark fragmentation functions as derived from the data with subtraction of the B-hadron decay charged particle momentum spectrum rescaled to the experimental decay multiplicity. (right) The corresponding results for $c\bar c$ events.}
\label{fig:decays}
\end{figure}

Figure~\ref{fig:decays} presents the results of the procedure outlined above. The Pythia8 MCEG predictions for the full distribution agree well with the data. The derived measurements of the fragmentation functions are obtained by subtracting the MCEG predictions for the momentum spectra of charged heavy hadron decay products from the data. 

\begin{figure}[H]
\begin{tabular}{cc}
\includegraphics[width=0.5\linewidth]{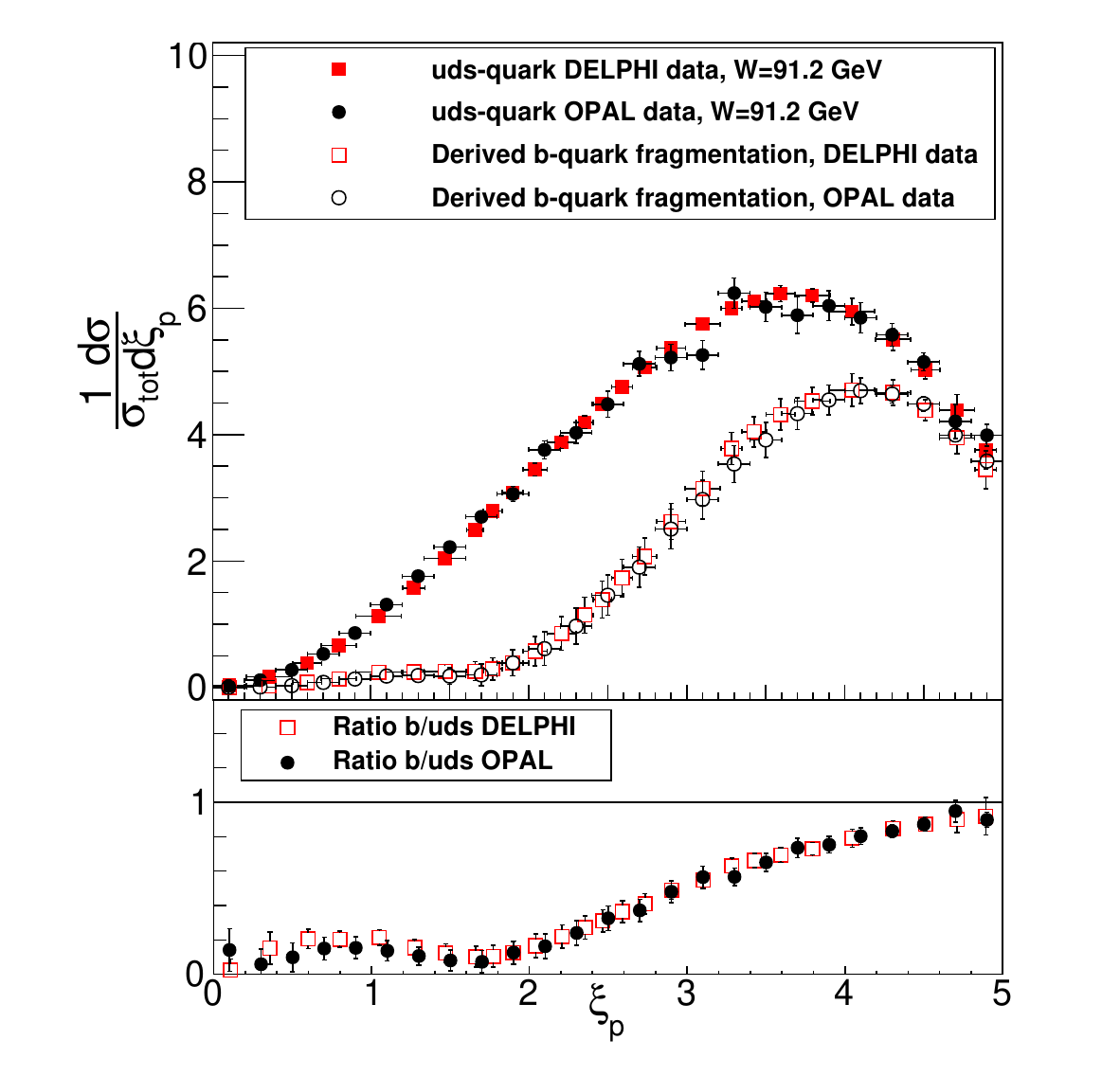} &
\includegraphics[width=0.475\linewidth]{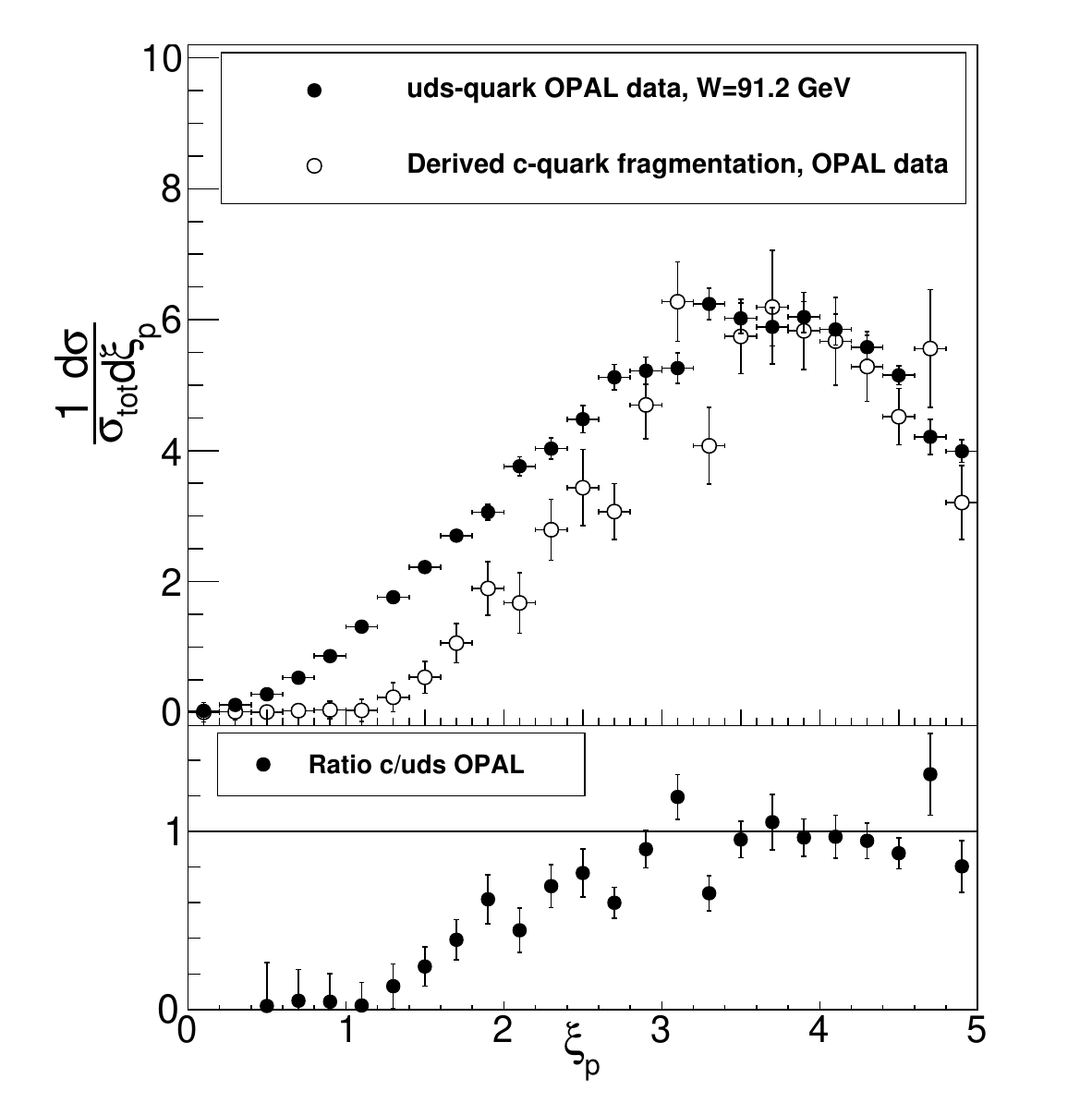} \\
\end{tabular}
\caption{(left) Momentum spectra for light quark jets in comparison with experimentally derived momentum spectra for b-quark jets. The lower panel
shows the bin-wise ratios of the heavy quark over the light quark momentum spectra. (right) The same plots for c-quark jets. Both for b- and c-quark jets the suppression of large momentum particles predicted by the dead cone effect is clearly visible.}
\label{fig:exp-deadcone}
\end{figure}

In order to study the dead cone effect in momentum space the results shown in fig~\ref{fig:decays} for the momentum spectra of charged particles are compared to the corresponding spectra from light quark jets. These have been measured by DELPHI and OPAL in the same analyses~\cite{DELPHI:1998cgx,OPAL:1998arz}. We derive bin-wise ratios of the momentum spectra in $\xi$ for heavy and light quark jets as shown in figure~\ref{fig:exp-deadcone}.

The ratios of momentum spectra measured with heavy quark jets over light quark jets show a suppression of up to a factor of about 10 for large momenta corresponding to small values of $\xi$. The question is now if this suppression is consistent with the predictions~\cite{Dokshitzer:1991fc,Dokshitzer:1991fd}
based on the description of the dead cone effect for momentum spectra in MLLA QCD. 

\begin{figure}[H]
\begin{tabular}{cc}
\includegraphics[width=0.45\linewidth]{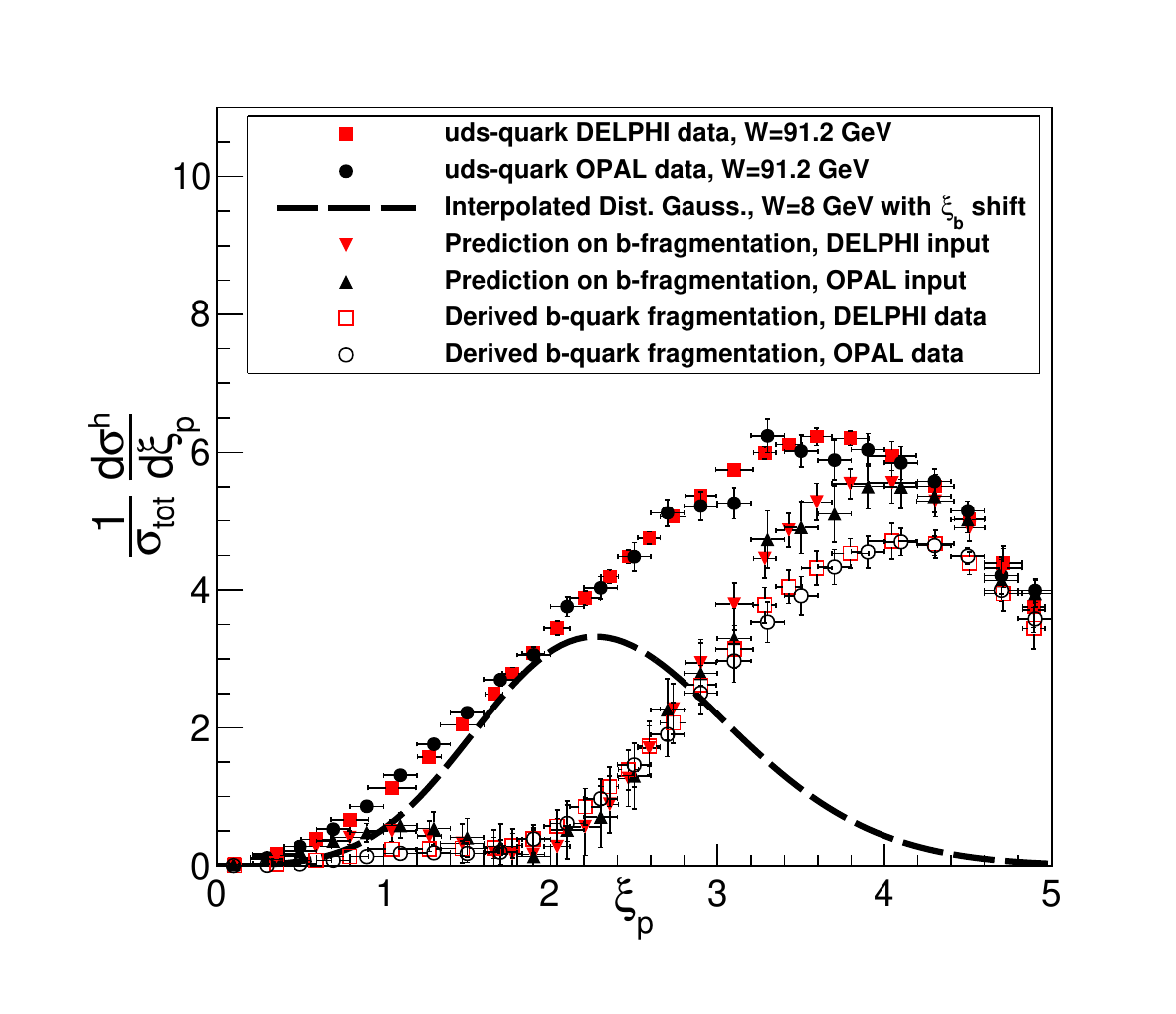} &
\includegraphics[width=0.475\linewidth]{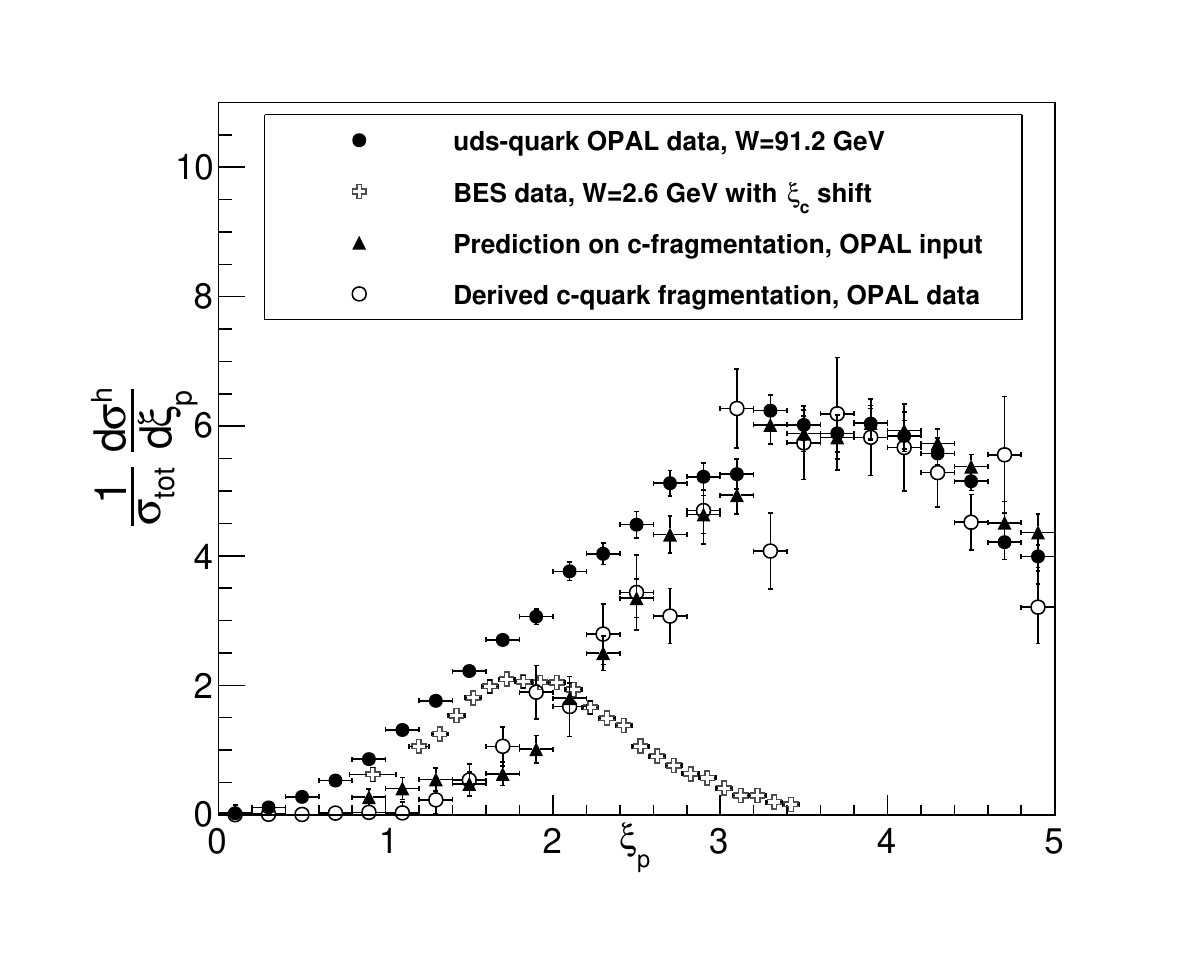} \\
\end{tabular}
\caption{(left) Momentum spectra for light (uds) quarks at 91.2~GeV and at $W_0=8.0$~GeV by interpolation between neighbouring energies, with correction for charm decays (see~\cite{Kluth:2023umf}). Subtracting these distributions as in equ.~\eqref{MLLAeqxi} yields the MLLA prediction for the b-quark momentum spectrum. Finally the b-quark momentum spectra derived from the data are shown. (right) corresponding results for c-quarks, with $W_0\sim 2.6$~GeV and using BES data~\cite{BES:2003xdf}.}
\label{fig:MLLA-comparison}
\end{figure}

An equation for the inclusive $x$-spectra has been presented which reproduces the predictions for multiplicities in MLLA~\cite{Dokshitzer:2005ri} after integration over $x$. This MLLA estimate has been reported in~\cite{Dokshitzer:1991fd} and is presented here in terms of $\xi$:
\begin{equation}
  \bar D_Q(\xi,W) = \bar D_q(\xi,W) - \bar D_q( \xi - \xi_Q,\sqrt{e}M_Q)\;\;,
\label{MLLAeqxi}
\end{equation}
with $\bar{D}_{Q,q}(\xi,W)=x D_{Q,q}(x,W)$ and $\xi_Q=\log(1/\langle x_Q\rangle)$. The mean momentum fraction ${\langle x_Q\rangle}$ of the primary heavy quark $Q$ is introduced which reduces the light particle energies to $x < {\langle x_Q\rangle}$. The shift of the $\xi$-spectrum by $\xi_Q$ corresponds to an MLLA correction of ${\cal O}(\sqrt{\alpha_s})$. The eq.~\eqref{MLLAeqxi} represents an approximation that does not work well at small $\xi$ since the shifted contribution $\bar D_q(\xi-\xi_Q,W_0)$ has to vanish for $\xi<\xi_Q$. This prediction implies that the momentum spectrum of charged particles in heavy quark $Q$ jets is connected with the difference of momentum spectra of light quarks $q$ evaluated at different energy scales $W$ and $W_0=\sqrt{e}M_Q$. In figure~\ref{fig:MLLA-comparison} we show results of determining the difference between light quark momentum spectra at scales $W$ and $W_0$ using data obtained at low energies for the subtraction term at scale $W_0$. The low energy data are interpolated in the case of b quark jets at $W_0=8$~GeV and are taken directly from BES data~\cite{BES:2003xdf} for c-quark jets at $W_0=2.6$~GeV. 

We find good agreement between the heavy quark momentum spectra derived from the data for b- and c-quark jets and the MLLA predictions based on light quark jet data. For b-quark jets there is an excess of the MLLA prediction over the data at large $\xi$. Such an excess of the MLLA prediction has been observed before in the analysis of charged particle multiplicities in b-quark jets~\cite{Dokshitzer:2005ri} and this excess is now seen to be correlated with large values of $\xi$. For c-quark jets there is good agreement over the whole range of $\xi$ within the uncertainties of the data. This comparison confirms that the observed strong suppression of charged particle production at large momenta in heavy quark jets w.r.t.\ light quark jets is indeed a direct consequence of the dead cone effect in QCD.

As a further consistency check we fit the central region $1.6 < \xi < 2.6$~GeV of the momentum spectrum in b-quark jets with the MLLA prediction in terms of light quark jet data and a free value of $W_0=\sqrt{e}M_Q$. The result is $W_0^{exp}=(7.5\pm0.5)$~GeV using the DELPHI and OPAL data. The fit result is consistent with the expectation $W_0=(8.0\pm0.2)$~GeV based on $M_b(M_b)=(4.85\pm0.15)$~GeV. There is a sensitivity of O(few\%) to the mass of the b-quark in the data for the dead cone suppression at large particle momenta from Z boson decays to heavy quarks in $e^+e^-$ annihilation. However, due to the limitations given by the residual model dependence of the data analysis (MCEG subtraction of heavy hadron decay products) and the MLLA QCD prediction this sensitivity is not competitive with other more advanced determinations of the b-quark mass. 

\section{Conclusions and Outlook}

In~\cite{Kluth:2023umf}, summarised here, the dead cone effect of suppressed radiation off heavy quarks in a cone around the flight direction, is confirmed by an analysis in momentum space by considering the momentum spectra of charged particles in heavy quark jets. The analysis uses data from the LEP experiments DELPHI and OPAL taken at cms energy $\sqrt{s}=M_Z$ on the Z boson peak. The measurements had to be corrected for the presence of charged heavy hadron decay products using predictions from the MCEG Pythia8. In comparison with momentum spectra in jets originating from light quarks in the same analyses a strong suppression of particles with large momentum fraction of up to a factor of about 10 is observed, establishing the dead cone effect for heavy quark jets with energies $E_{jet}\simeq 45$~GeV. The analysis is model independent, except for the MCEG based subtraction of heavy hadron decay products, and can be interpreted directly with MLLA QCD predictions. 

The dead cone effect in momentum space could have interesting consequences for heavy flavour tagging of jets in particle physics experiments. Traditional tagging algorithms consider the partial observation of the corresponding heavy hadron decay via e.g.\ the presence of secondary vertices, tracks with large impact parameter, or large momentum leptons. In machine learning methods using deep neural networks together with presenting all jet constituents (see e.g.~\cite{Qu:2019gqs,ATLAS:2022rkn}) the dead cone effect in momentum space, i.e.\ the strong suppression of large momentum particles in heavy flavour jets, is expected to contribute to the discrimination between heavy and light flavour jets. It could be useful in this context to study the $\xi=\log(1/x)$ scaled momentum spectra in heavy and light flavour jets at the LHC and compare measurements with MCEG predictions, because any mismodeling of the momentum spectra by MCEG programs could bias machine learning heavy flavour tagging algorithms trained on MCEG event samples. 

It will be interesting and challenging to translate the present study to an analysis of the dead cone effect in top quark production at the LHC.

%%%%%%%%%%%%%%%%%%%%%%%%%%%%%%%%%%%%%%%%%%

\authorcontributions{All authors have contributed equally to all aspects of this paper.}

\funding{This research received no external funding.}

\conflictsofinterest{The authors declare no conflict of interest.}

%%%%%%%%%%%%%%%%%%%%%%%%%%%%%%%%%%%%%%%%%%
\begin{adjustwidth}{-\extralength}{0cm}
%\printendnotes[custom] % Un-comment to print a list of endnotes

\reftitle{References}

% Please provide either the correct journal abbreviation (e.g. according to the “List of Title Word Abbreviations” http://www.issn.org/services/online-services/access-to-the-ltwa/) or the full name of the journal.
% Citations and References in Supplementary files are permitted provided that they also appear in the reference list here. 

%=====================================
% References, variant A: external bibliography
%=====================================
%\bibliography{hbpdeadcone}

%%%%%%%%%%%%%%%%%%%%%%%%%%%%%%%%%%%%%%%%%%

\PublishersNote{}

\end{adjustwidth}

\end{document}